\documentstyle{article}
\begin{document}
\hspace{1cm}  \vspace{1cm}

\begin{center}
	 {\Large Network of  Hydrogen Bonds \vspace{5mm}  \\
							 as a Medium for  DNA Interaction in Solvents }
							 \vspace{1cm} \\
							 {\large V.L. Golo}$^{\dagger}$,
							 {\large E.I. Kats}$^{\dagger \dagger}$, and
							 {\large Yu.M. Yevdokimov}$^{\dagger \dagger \dagger}$
							 \vspace{1cm}\\
							 $^{\dagger}$ Department of Mechanics and Mathematics \\
													Moscow University, Moscow 119899, Russia \vspace{5mm} \\
							 $^{\dagger \dagger}$ Landau Institute for Theoretical Physics\\
													 Russian Academy of Sciences \\
													 Moscow, Russia \vspace{5mm}	\\
							 $^{\dagger \dagger \dagger}$ Institute of
													 Molecular Biology \\
													 Russian Academy of Sciences \\
													 Moscow, Russia
\end{center}

\noindent
{\Large Abstract} \vspace{5mm}

\noindent
We suggest that the DNA molecules could form the cholesteric phase
owing to an interaction mediated by
the network  of the hydrogen bonds (H-network)
in the solvent.  The model admits of the dependence of the optical
activity of the solution on the concentration of the PEG, and the
change in the sense of the cholesteric twist due to the intercalation by
the daunomicyn. Using the experimental data for the cholesteric phase
of the DNA dispersion, we obtain a rough estimate
for the energy given by our model, and show that
it should be taken into account as well as the energy due to the steric
repulsion, van der Waals, and electrostatic forces, generally used for
studying the DNA molecules.
The elastic constant of the H-network generating
the interaction between the DNA molecules is determined by the energy due
to the proton's vibration in the hydrogen bonds.

\pagebreak

\noindent 1.
It is generally accepted that the formation of the cholesteric phases
of the DNA dispersions is intimately related to the interplay among
the basic properties of the DNA and the solvent, but so far the nature
of the interaction, in the dispersions,  between the molecules of DNA
has remained obscure. In this communication we wish to suggest a model
based on the structure of the network of the hydrogen bonds in water,
which could explain the cholesteric transitions in the DNA dispersions.
The main point is that the conformation of the network of the hydrogen
bonds is influenced by the presence of the DNA molecules, so that the
energy of the network is affected, and therefore one may expect
a feed-back effect resulting in an interaction between the
DNA molecules themselves.

\noindent
It is worth noting that the hydrogen bonds, even though weak, $5\, kcal/mol$,
that is much less than the covalent ones, are still larger than the
van der Waals ones, and strong enough to hold the water molecules in a
rigid structure in solid water (ice), \cite{Pauling}.
The structures of the water and the ice are similar,
and either of them is determined by the network of the hydrogen bonds,
\cite{BF}, even though in liquid water some of the
hydrogen bonds are broken; their number depending on temperature.
The whole may be visualized as the fisherman's net; the knots being atoms
of the oxygen and the threads corresponding to the hydrogen bonds.
The conformation of the net and the distribution of the protons are
determined by Bernal-Fowler's rules, i.e. there are two protons in  the
neighbourhood of an atom of the oxygen, and one proton is assigned
to each hydrogen bond. We should allow also for defects; some of
the threads (H-bonds) being ruptured owing to thermal fluctuations.

\noindent
The conformation of the network of the
hydrogen bonds in the DNA dispersions has certain properties that make it
quite different from the network of the hydrogen bonds in the bulk solvent,
so that the interaction between the DNA molecules at a distance
less than $50 \AA$ may result in new phenomena.
The reason for this is twofold.

\noindent
First, the quantum mechanical computations performed
in a number of papers, \cite{Neidle}, \cite{Saenger}, show that the water
molecules close to the DNA, i.e. at a distance of a few $\AA$,
form a structure that follows the geometry of the molecule of DNA
so that the set of the H-bonds is strictly determined by
the DNA conformation, whereas in the bulk,
at a distance of $10 \AA$ to $50 \AA$,
the influence of the DNA  is considerably weakend.
Starting from classical crystallogrphic arguments,
N.A. Buljenkov, \cite{Bul}, arrived at a conclusion
that the main factor in the making of a conformation for the ensemble
of the water and the DNA molecules, is the H-bonds.
According to \cite{Bul}, inside one of the grooves of
the DNA the water molecules form a kind of helix,
whereas outiside the remaining water
molecules are arranged in a hierarchical fashion,
or,  putting it in more quantitative terms,
they should form a fractal structure
according to a recursive algorithm, \cite{Bul}.

\noindent
Second, the numerical simulation by Paola Gallo, \cite{Paola},
indicates that there are two distinct sets of water molecules confined in
a silica-pore, i.e. those sufficiently close to the wall, and those
outside in the bulk.  The molecules close to the surface of the pore
tend to form H-bonds with the atoms of the substrate so that the H-network
close to the walls is strongly distorted, whereas the water
in the bulk region, a few layers outside of the surface,
shows the boson peak anomally, which is generally considered as
a precursor of the glass transition, \cite{Paola}.
The size of the silica pore is similar
to that of the inter-molecular distance of the DNA dispersions,
so that allowing for the difference in the potentials of surface interaction,
one may suggest that the water molecules confined in the DNA dispersion may
exhibit a similar behaviour.  In this respect it is worthwhile to recall
papers \cite{Par1}, \cite{Par2}, that claim the importance of
what they call hydration forces, which dominate at close separations
between the DNA molecules, so that the resultant interaction  resemble
neither the pure electrostatic nor the steric one.

\noindent
There are experimental data suggesting that
there be a need for taking into account the part
played by the inter-molecular medium of the DNA dispersions.
Indeed, the transitions from the cholesteric to the nematic phases,
which are observed in the DNA dispersions  ,
have a number of features that
run considerable difficulties in explaining
within the conventional framework of electrostatic
and van der Waals forces, (see paper \cite{kk} in which the problem is
studied within the framework of the electrostatic interaction between
the molecules of the DNA).

\noindent
Let us recall the most prominent experimental facts.

\noindent
 (i)
The liquid crystalline phases formed from the
double-stranded DNA, exhibit a negative band in the CD spectra,
whereas the liquid crystalline-phases and  dispersions formed from the RNA,
under the same conditions, have a positive band in their CD spectra,
\cite{Y1}. The positive (resp. negative) sense of the CD band is
alleged to correspond to the left (resp. right) twist of cholesteric.
Under the experimental conditions of \cite{Y1}, the difference
between the right-helix of the B-form of the DNA and
the right double-helix form of the RNA
appears to be not sufficiently significant, there is
only the change of thymine for uracil,
to cause a substantial spatial
effect solely because of the electrostatical and the van der Waals forces.
In fact, one should expect only a tiny modification of
the inter-molecular interaction, and not the reconstruction
of the total structure.

\noindent
 (ii)
In paper \cite{Y1} it is reported that nucleic acid molecules form
the cholesteric phases in the presence of the poly(ethylene glycol), (PEG),
only at a certain osmotic pressure of the PEG.
The osmotic pressure can be controlled through the variation of the
concentration of PEG in the solvent; to a certain
extent, it is similar to the classical experiment with a semipermeable piston
that exerts a pressure equal to the osmotic pressure of the PEG solution.
It is generally considered to be an efficient means for estimating the
repulsion interaction between the DNA helices in solution, \cite{Par1},
\cite{Par2}.
The results of paper \cite{Y1} are illustrated in Fig.1; it appears that
there should be a change in the state of the dispersion.
There is an obvious correlation between
the character of the osmotic curves  and the optical activity of the
solution, which should correspond to the formation of the cholesteric phase.
But the osmotic effect is due to the action of the hydration force
associated with the removal of the water
molecules from the DNA environment. Therefore, the force generates
the principal interaction between DNA molecules,
rather than the van der Waals or the electrostatic one (cf. \cite{Par2}).

\noindent
 (iii)
The CD spectra of the liquid crystalline dispersions
formed from the right handed-helices of the B-DNA
has the important characteristic, the intensive negative band, whereas for
the case of the A-DNA, the corresponding band is positive.  One may
change the CD-spectrum of the DNA dispersion by dyes, so that
two bands instead of one appear. Depending on the dye
one employs, the two bands may vary.
If the dye is intercalated between the
DNA base-pairs, the two bands are related to (1) the absorption region
of the DNA nitrogen bases, and  (2) the absorption region of the compound
itself, respectively.  If, as is the case with the daunomycin, a part
of the compound lies "outside", i.e. in one of the grooves of the
DNA, there appear two positive bands in the CD spectrum of the molecule.
It is alleged that the two positive bands are
the signature of the right-handed
twist of the cholesteric helix formed by the DNA molecules.
Thus, the sense of the twist is changed by the daunomycin.
But the daunomycin can be eluted by sodium dodecylsulfate (SDS),
so that the positive band related to
the daunomycin disappears; nonetheless the positive band in the
absorption band of the DNA still remains.  One may suggest that
the right-handed twist of the cholesteric structure of the DNA dispersion remains
as well. Thus, it is possible to obtain the right-handed twist cholesteric
structure, besides the usual left one, for the same B-form of the DNA.
One may suggest that if the cholesteric twist were determined by
the van der Waals and the electrostatic forces, as well as the form of the
DNA, the co-existence of these two metastable states should be unlikely.

\noindent
 (iv)
According to \cite{Y1}, even small local defects in the structure
of the DNA molecules may cause the change in the cholesterical structure.
It is important that defects of the surface of the DNA molecule result in
the formation of a helical structure in which the sense of the cholesteric
twist differs from the sense of the twist of the DNA cholesteric, and the
intense band in the CD spectrum must tend to zero as the number of the
defects increases. It is possible that the defects affect the neighborhood
of a DNA molecule, thus preventing the formation of the cholesteric packing;
the steric, van der Waals, and electrostatic effects not being involved,
\cite{Y0}.  \vspace{1cm}

\noindent
 2.
As was explained above, in the presence of the molecules of DNA, e.g.
in the dispersions, the network of the hydrogen bonds  has different shapes
in the regions close
to the DNA molecules and in those of the bulk, a few layers of the water
molecules away from them. One may imagine that close to
the DNA molecules, the net looks like being rolled round a pole;
it is spreading more leisurely away from the DNA molecules.

\noindent
The key problem is to find an estimate for the elastic
properties of the system. In fact,
the differences in energy between microscopical
states of the network of the hydrogen bonds are small,
and therefore the configuration of the
system is determined by entropy. Therefore, the H-network
may be considered, from the point of view of its elastic properties,
as a system of the "athermal" type, \cite{Kantor}. Presently, the
information about their elastic properties is scarce, so that
one may expect that the study of the cholesteric phases of
the DNA dispersions may provide a new insight into the matter
(see below). We shall assume that the network of the H-bonds in solvent
has a certain amount of elasticity, and in fact enough to account
for the interaction of the DNA molecules.
At the same time the system may also have some elastic energy of the usual
thermal type; presumably, not much, depending on the elasticity of
the H-bonds, i.e. their being extensible.
The total energy comprized of the elastic and the entropy term,
under constraints imposed by the boundary effects,
should result in an effective iteraction between the DNA molecules,
through the medium of the network of the hydrogen bonds. It is to be
noted that we have neglected the van der Waals and the electrostatic forces,
usually suggested to be of primary importance, on the ground of the
arguments given in the preceeding section. Their effect could still
be important in the system's choosing its orientation as regards
the cholesteric twist (see below), even though the determining factor
should be the interaction mediated through the H-bonds.

\noindent
The picture given above implies that there is a neighbourhood of
the DNA molecule in the solvent in which the conformation of the set
of the molecules of water is strongly influenced by the geometry
of the DNA helix. This assumption is in full agreement with
the theoretical and experimental results reported in  the preceeding
section. In this neighbourhood, which spreads out
at a distance of approximately $10 \AA$,
the conformation of the H-bonds follows the grooves of the DNA
and has the shape resembling that of the helix. In this respect it
differs considerably from that in the bulk, (compare the previous
section, \cite{Saenger},  \cite{Paola} ).

\noindent
According to the numerical simulation of \cite{Neidle}
the water molecules in the major groove
form a regular framework comprized of hydrogen bonded pentagons that
lie in the same plane as the base-pairs; in contrast, the water molecules
in the minor groove form only a monolayer.
Consequently, the conformation of the network of the H-bonds close to
a molecule of the DNA could be compared, at least qualitativeley,
to a helical, or winding, dislocation. The latter has the specific
feature of being a kind of singularity in the fabric of  the bulk network
of the H-bonds, and in this sense one may consider it as a topological
dislocation.

\noindent
If we assume that the elastic, perhaps totally "athermal", or entropy
generated, interaction between the dislocations of the network of the H-bonds
is at work, we may suggest as well, within the  usual framework
of elasticity theory, that two parallel dislocations of the type
interact according
to the usual logarithmic law, i.e. the energy per unit length
of the interaction of two
parallel dislocations is given by the equation
\begin{equation}
	 E_{disl} = - \frac{\epsilon\,	e_1  e_2}{2}\,  \ln r  \label{log}
\end{equation}
in which $\epsilon$ is the elastic modulii, $ e_1,  e_2$ are the "charges"
determined by the winding of the net at the cores of the dislocations,
that is the molecules of the DNA, and $r$  is the distance between the
dislocations. The interaction is the repulsive one. According to the
arguments given above, Eq.(\ref{log}) gives the energy of the interaction
of the two parallel molecules of the DNA, through the medium of
the network of the H-bonds.

\noindent
Next, let us consider two molecules of the DNA that are not parallel.
Since the cholesteric phases in DNA dispersions are studied for
segments of DNA of sufficiently low molecular weight,
we may suppose that the molecules are of a length $l$,  smaller than the
persistent one, and we may visualize them as segments of two skew lines.
For the sake of simplicity, and allowing only for a rough
estimate of the energy of their interaction, we suppose as well that
their centres be on the same line, at a distance $d$;
the segments being at an angle $\phi$ to each other.
We shall use the planar approximation for the interaction,
that is we assume that only bits of the molecules lying
in the same planes do interact in accord with Eq.(\ref{log}),
so that the corresponding energy reads
\begin{equation}
		 F_{bits} = - \frac{\epsilon}{2}\, \ln
										\left[
														 d^2 + 4\, \left (
																				\frac{x}{l}
																		 \right )^2
																	\sin^2 (\phi / 2)
										 \right ]
									 + \epsilon \ln a
											\label{bits}
\end{equation}
where $x$ is the position of the bits on the line of the molecule,
and $a$ is the radius of the core of the dislocation, i.e. the radius
of the DNA molecule.
To derive the energy of the total interaction of a pair of
the DNA molecules, we shall take the integral of the function defined by
Eq.(\ref{bits})
\begin{equation}
	 U_{disl}  = \int_{-l/2}^{+l/2} F_{bits}\, dx \label{disl}
\end{equation}

\noindent
The energy given by Eq.(\ref{disl}) is only a part of the total expression
for the interaction energy of the DNA molecules in solvent. It is
necessary to allow for the isotropic pair interation, $U_0$,
the conventional
energy due to the van der Waals, electrostatic and steric forces, $U_1$,
the anisotropic contribution related to the chiral state, $U_2$, so that
the total expression has the form
$$
			 W_{\phi} = U_0 + U_1 + U_2 + U_{disl}
$$
It is important that the term $U_0$ does not depend on the angle $\phi$
describing the mutual orientation of the molecules.
The twist angle, $\phi$, is small. Therefore, the term $U_1$ is
of the second order in $\phi$, and $U_2$ of the third,
\begin{equation}
	U_1 = A\, \phi^2, \qquad  U_2 = B\, \phi^3  \label{U_12}
\end{equation}
In what follows we shall assume that the contributions due to
$U_1$, which takes into account the combined influence of the steric,
van der Waals, and electrostatic forces,  is comparable with  $U_{disl}$,
whereas $U_2$ is a small correction.
It is also necessary to
take into account the energy due to the proton transfer,
but within the approximation used in this paper it does not depend on
the cholesteric twist, $\phi$, and we may postpone considering the effect
until Section 3.

\noindent
On expanding $U_{disl}$ to within the fourth order in $\phi$,
and neglecting for a while the contribution due to $U_2$,
we obtain the following expression for $W_{\phi}$
\begin{equation}
 W_{\phi}	=  - \lambda\, \ln \left (\frac{d}{a} \right )
						 + \left [ A - \frac{\lambda}{3}
											\left ( \frac{l}{d} \right )^2
							 \right ] \phi^2 \\ 
						 + \frac{\lambda}{2} \left ( \frac{l}{d}  \right )^4 \phi^4
             	\label{total}
\nonumber
\end{equation}
in which $  \lambda = l\, \epsilon $.
It is worth noting that for the experimental data, i.e. $d$ within
$30\, \AA$ to $50\, \AA$, $l$ of the size $500\, \AA$, and $\phi$ of
the order $0.01\, rad$, we have the small parameter $\kappa = l\phi/d$,
and Eq.(\ref{total}) is an expansion of $W_{\phi}$ in $\kappa$.
The minimization of $W_{\phi}$ gives the value of the cholesteric twist
angle
\begin{equation}
	\phi^2_*  = -\, \frac{d^2}{l^4} \left ( \frac{A}{\lambda}d^2
																		 - \frac{l^2}{3}
															\right )
									 \label{twist}
\end{equation}
At first sight, the cholesteric phase appears to exist
for all intermolecular distances $ d \le d_0 = l\, \sqrt{(3 A / \lambda)}$.
In fact, the situation is different.
For one thing, for small values of $d$, say less $20\, \AA $,
the asumption that the H-network be soft enough, breaks down, for the
hydrogen bonds form a rigid structure due to the surface effects of the
DNA molecules, and in this region of $d$
the nematic phase is observed, \cite{Par1}, \cite{Par2}, \cite{Y1}.
For another, the value of $d_0$  is determined by
the relative sizes of the elastic constants $A$ and $\lambda$.
According to  \cite{Par1}, \cite{Par2}, \cite{Y1}, the cholesteric phase
is observed for the values $d \approx 30\, \AA$ and $l \approx 500\, \AA$,
therefore, we may conclude that $A$ could be
less than $\lambda$ at least by an order of magnitude,
so that $d_0$ may be of the same order as $l$.
Thus, Eq.(\ref{twist}) describes the cholesteric twist $\phi_*$,
within the range of $30\, \AA$ to $50\, \AA$ for $d$.
Our discussion is very crude, and qualitative, so that we may only
infer from Eq.(\ref{twist}) that the twist angle, $\phi_*$, increases
with $d$ in the region indicated above, and in this respect our result
agrees with the formular for $\phi_*$
$$
	\phi_* \propto \frac{\sqrt{d}}{l}
$$
obtained in paper \cite{kk} within the framework of
the electrostatic approach. But the region of $d$ is too small to make
any asymptotic conclusions, especially that a significant
dependence of $\phi_*$ on the spatial separation $d$
has not been observed, so far. It should be noted, as well,
that paper \cite{kk} neglects the conclusions of
Parsegian et al, \cite{Par1}, \cite{Par2},
that the hydration forces be of primary importance, and does not
consider the PEG experiments.

\noindent
It is important that below the threshold we have the two values for
the twist angle, $\phi_+$ and $\phi_-$, owing to the energy $U_2$,
(see Eq.(\ref{U_12})), which takes off the degeneracy in $\pm \phi_*$.
Consequently, we may expect the existence of two metastable
cholesteric structures, above the nematic threshold. This circumstance may
have a bearing upon the change of the CD spectrum due to the daunomycin,
described above. We suggest that under the action of the latter,
in a layer of thickness of approximately $5\, \AA$,
the conformation of the surface is deformed so that
the structure of the hydrogen bonds in the whole volume is changed in
such a way that the system comes down to the minimum different from that
it would occupy in the absence of the daunomycin.
 
\noindent
Conformational changes in the DNA molecules could be a cause for the
reconstruction of the system of the hydrogen bonds in the bulk of the
solvent, and therefore result in the formation of various
cholesteric structures. In this respect,
small local defects mentioned above,
as well as the conformation structures of
the double-helix of the B-DNA and the double helix form
of the A-RNA, may bring about changes in the network of the hydrogen bonds
that could result in different liquid crystalline structures. Similarly,
the PEG solutions may provide conditions for the formation of
different networks of the hydrogen bonds that could generate effective
interactions between the DNA molecules
that may result in liquid crystalline structures,
The phenomena are difficult to explain within the framework of the
theory using only electrostatic and van der Waals forces.

\noindent
So far we have considered the network of the
H-bonds generally with respect to its conformation, that is
the knots corresponding to the positions of the atoms of oxygen
and the threads to the H-bonds. It is important that stringent topological
conditions should be imposed on the network owing to the requirements
due to the structure of ice, \cite{Pauling}, i.e.
among $16$ possible combinations
of the protons round an atom of the oxygen only $6$ are allowed,
so that the residual entropy of ice turns out to be equal to
$S_0 = Nk\, \ln (3/2)$.
The topological constraints are also essential for the proton dynamics of
the H-bonds. In this respect it is worth noting the phenomenon reported
in paper \cite{Saenger} that the protons could perform cooperative or
concerted motions (flip-flop) by jumping from one state to another.
Saenger et al, \cite{Saenger}, claim that one could find chains of
the water hydroxyl groups
$O - H \cdot \cdot \cdot O - H \cdot \cdot \cdot O - H $
in which the protons would oscillate clockwise and anti-clockwise in a
flip-flop manner. Obviously, the choice of the circles allowing for the
flip-flop motion is determined by the topology
of the network of the H-bonds.  \vspace{5mm}

\noindent
 3.
It is instructive to make a rough estimate of the effect produced by
the proton's vibration. One can visualize it as a motion
parallel to the valence bond $O - H \cdot \cdot \cdot$,
in a cell of the order of $\sigma$;
$\sigma$ being  the size of the hydrogen bond.
In fact, the proton moves in the field of the double-well potential
inside the hydrogen bond, with a characterisitic frequency, $\nu_s$,
which is of the order $3000\, cm^{-1}$; the corresponding energy
being $\hbar\, \omega_s$ per H-bond, \cite{Pauling}. Here we take into
account only the lowest frequency, and neglect the obertones.
The number of these cells per
area of the cross section of the volume in the DNA dispersion depends
on the separation $d$ of the molecules. The value of $\sigma$
could not change appreciably with $d$.
Consequently, the number of the hydrogen bonds
per pair of the DNA molecules
involved in the conformation of the cholesteric structure should be
proportional to
\begin{equation}
 N_H = 		\frac{l}{\sigma} \left	(	\frac{d}{\sigma} \right )^2 \label{NH}
\end{equation}
so that  the energy due to the proton vibration may be estimated at
\begin{equation}
	 U_P = \frac{C}{2}\, \frac{l}{\sigma}
	 \left ( \frac{d}{\sigma} \right )^2\, \hbar\, \omega_s
	 \label{UP}
\end{equation}
where $C$ is a constant, presumably of order $1$.  It is obviously
similar to the zero point energy, familiar in the theory of
the so-called Casimir effect.

\noindent
The zero point energy may play a significant
role in determining macroscopical properties of a system. For example,
its large value for the helium atoms is the reason
why the volume of liquid helium at zero pressure is almost three times as
large as the volume it would have had in its absence, \cite{London}.
In the situation of the network of the hydrogen bonds, it is important
that the energy of the proton's vibration,
$\hbar \omega_s \approx 6\, 10^{-13}\, erg$, is approximately $16\%$ of the
energy of the hydrogen bond, $5\, kcal/mol$, or $3.46\, 10^{-12}\, erg$.
Hence, one may expect that the interplay among the elastic energy of the
H-network and the vibrational energy of the protons should
result in a meaningful structure.
In fact, it does, and the expressions for $W_{\phi}$ and $U_P$
having opposite signs insures that the formation of the dislocations in
the H-network allows the system to lower its total energy. We may cast
these arguments in a more quantitative form in the following way.

\noindent
Using the expression for the energy $W_{\phi}$
we may cast the energy of the interaction of
the two molecules in the form
$$
	E_H = W_{\phi}	+ U_P
$$
and on minimizing  $E_H$ we obtain Eq.(\ref{twist}) and the second
equation,  $\partial E / \partial d =  0$,
which can be cast in the form
\begin{equation}
	 \lambda \approx C\, \frac{l}{\sigma} \left ( \frac{d}{\sigma}
																				\right )^2 \hbar \omega_s
																				\label{ll}
\end{equation}
From Eqs.(\ref{twist}-\ref{ll}), we may derive an estimate for
the elastic constant $\lambda$. In fact, since
$\nu_s \approx 3000 cm^{-1}$, and  $d$ and $l$ of $50\, \AA$ and $500\, \AA$,
respectively, (see \cite{Y1}), we get $\lambda \approx 10^{-8}\, erg $.
According to the arguments given above, $A$ is less
than $\lambda$, by at least an order of magnitude.

\noindent
At this point it is worthwhile to draw attention to the part played by
the PEG. The latter is alleged to influence the hydration forces, i.e.
to diminish the number of the H-bonds involved in the interaction of
the DNA molecules. Therefore, the presence of the PEG should reduce
the value of $N_H$ given by Eq.(\ref{NH}), and the constant $C$
should decrease as well. Consequently, in accord with Eq.(\ref{ll}),
the elastic constant $\lambda$ should decrease and therefore the
twist angle $\phi_*$ should diminish. But, the decrease in $\phi_*$
has not been observed.  This is, perhaps, due to the circumstance
that $A$ may significantly depend on $d$.

\noindent
 4.
Our hypothesis as to the nature of the interaction of the DNA molecules
that leads to the formation of the cholesteric phase, is in qualitative
agreement with exprimental data; it accomodates
\begin{itemize}
 \item the strong dependence
	 of the formation of the cholesteric structure and their properties,
	 on the conformational properties of the nucleic acids;
 \item the predominance of the hydration forces demonstrated with the
	experiments using the PEG solutions;
 \item the change in the sense of the cholesteric twist due to the
	intercalation by the daunomycin;
 \item the large effects caused by small defects
	of the molecular structure, and the different senses of the twist
	angles for the  DNA  and the RNA-molecules.
\end{itemize}
We have not found the isotope effect, and in fact so far it has not been
observed.

\noindent
But it is worth noting that the concept of the network of the H-bonds
as a medium for the interaction between molecules of nucleic acids in
dispersions, makes for studying the water itself. Indeed,
according to our model, one could employ
the DNA dispersions as a valuable probe into the structure of water,
and in this respect it is very promising that we have obtained
the reasonable elastic characteristics of the network of the hydrogen bonds.

\noindent
The authors are thankful to N.A. Buljenkov, V.I. Ivanov, and G.G. Malenkov,
for the useful discussions.

\pagebreak

\pagebreak

\centerline{Figure Captions}

\noindent
Fig.1  \\
The dependence of  $\log \Pi$ on average distance between nucleic acid
molecular axes $(D,\, \AA)$. \\
(1) DNA;  \\
(2) RNA. \vspace{5mm}

\noindent
Fig.2

\noindent
A. \\
The CD spectra of the liquid-crystalline dispersion
(curve 1) treated by DAU (curves 2,3).  \\
$C_{DNA}$ --- $28.2\, \mu g / ml $ ;\\
$C_{PEG}$ --- $150\, \mu g / ml $ ;\\
$(0.3 M\, NaCl + 0.005\, Na - phosphate \quad    buffer)$ ; \\
Curve 1 --- $C_{DAU} = 0$; \\
Curve 2 --- $C_{DAU} = 6.8 \cdot 10^{-6}\, M$; \\
Curve 3 --- $C_{DAU} = 27.2 \cdot 10^{-6}\, M$; \\
$\Delta\, A$ in $mm$; $1\, mm$ --- $10^{-5}$ opt. units.

\noindent
B. \\
The CD spectra of the liquid-crystalline dispersions formed by
(DNA-DAU) complexes.  \\
$C_{DNA}$ --- $28.2\, \mu g / ml$ ; \\
$C_{PEG}$ --- $150\, mg / ml$ ; \\
$(0.3\, M\, NaCl + 0.005\, Na-phosphate \quad   buffer)$ ; \\
Curve 1 --- $C_{DAU} = 13.6 \cdot 10^{-6}\, M$ ; \\
Curve 2 --- $C_{DAU} = 37.6 \cdot 10^{-6}\, M$ ; \\
$\Delta\, A$ in $mm$; $1\, mm$ --- $10^{-5}$ opt. units.


\begin{thebibliography}{99}
	\bibitem{Pauling} L. Pauling, The Nature of Chemical Bond,
			 (Cornell University Press, 1960).
	\bibitem{BF} J.D. Bernal and R.H. Fowler,
			 A theory of water and ionic solution,
			 J.Chem.Phys. vol.1, 515 (1933).
	\bibitem{Neidle} S. Neidle, H.M. Berman, and H.S. Shieh,
			 Highly structured water network in crystals of
			 a deoxydinucleoside-drug complex,
			 Nature vol.288, 129, 13 November 1980.
	\bibitem{Saenger} W. Saenger, Ch. Betzel, B. Hingerty, and G.M. Brown,
			 Flip-flop hydrogen bonding in a partially disordered system,
			 Nature vol.296, 581, 8 April 1982.
	\bibitem{Bul} N.A. Buljenkov,
			 The role of hydration in the organization of biosystems,
			 Biophysics, vol.36, 181 (1991).
	\bibitem{Paola} Paola Gallo, Single Particle Slow Dynamics of
			 Confined Water, cond-mat/0003027, (2000).
	\bibitem{Par1} D.C. Rau and V.A. Parsegian,
			 Direct measurement of temperature-dependent
			 solvation forces between DNA double helices,
			 Biophys.J., vol.61, 260 (1992)	.
	\bibitem{Par2} R. Podgornik, D.C. Rau, and V.A. Parsegian,
			 Parametrization of direct and soft steric-undulatory forces between
			 DNA double helical polyelectrolytes in solutions of several different
			 anions and cations,
			 Biophys. J., vol.66, 962 (1994).
	\bibitem{kk} A.A. Kornyshev and S. Leikin,
			 Twist in chiral interaction between biological helices,
			 Phys.Rev.Lett., vol. 84, 2537 (2000).
	\bibitem{Y0} Yu.M. Yevdokimov, V.I. Salyanov, A.T. Dembo, and F. Spiner,
			 The "recognition" of DNA molecules and their packing
			 in liquid crystals,
			 Sensory Systems, vol.13, 151, 1999.
	\bibitem{Y1} Yu.M. Yevdokimov, S.G. Skuridin, S.V. Semenov, V.I. Salyanov,
			 and G.B. Lortkipanidze,
			 Stability of optical properties of the DNA cholesteric liquid
			 crystalline dispersions,
			 Biophysics, vol.43, 240 (1998).
	\bibitem{Kantor} O. Farago and Y. Kantor,
			 The elastic behaviour of entropic "Fisherman's Net",
			 cond-mat/0004276, (2000).
	\bibitem{London} F. Simon, Behaviour of condensed helium near
		absolute zero, Nature, vol. 133, 460, 529 (1934).
\end{thebibliography}
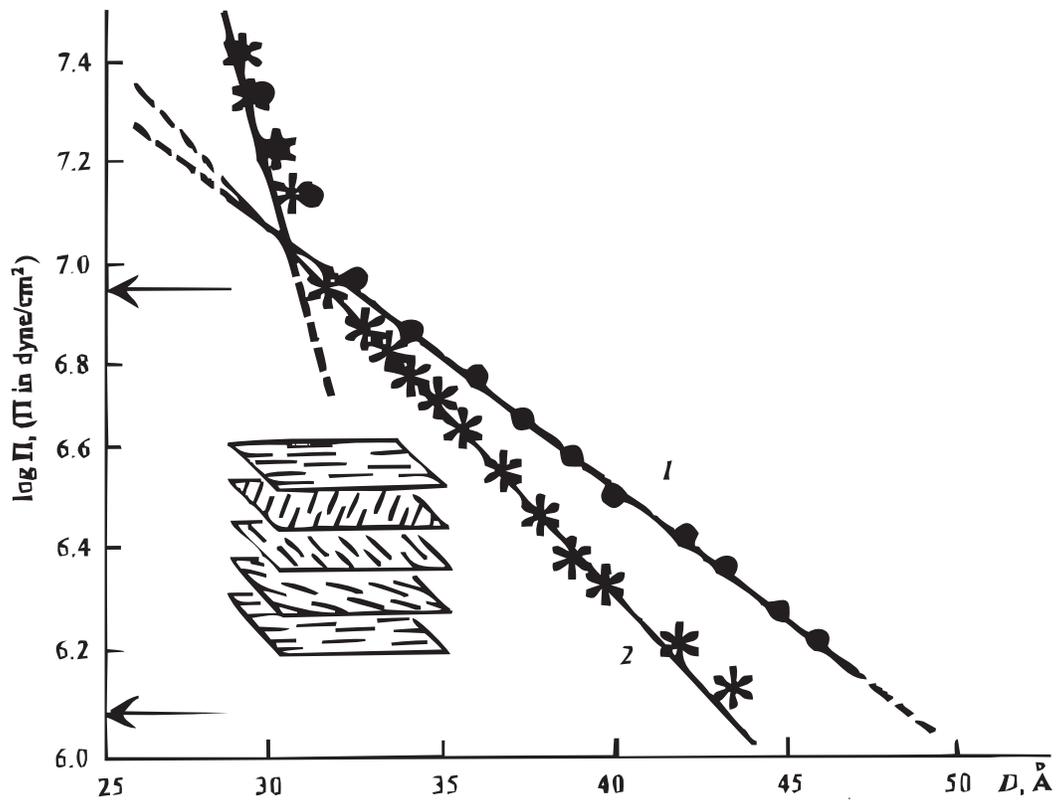
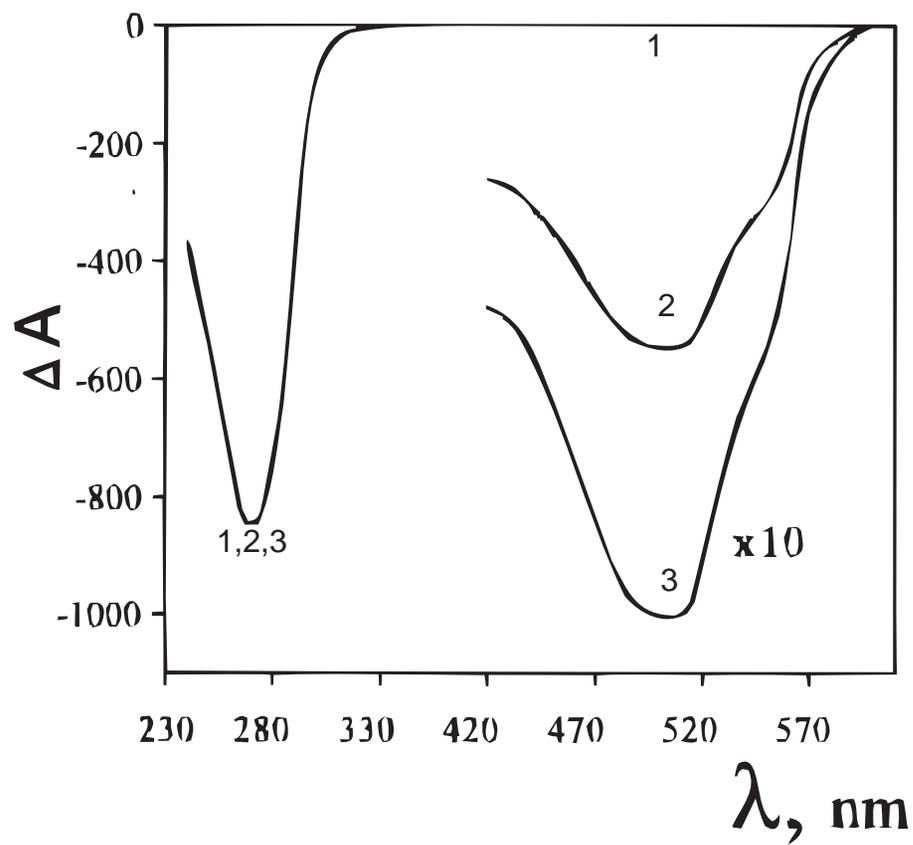
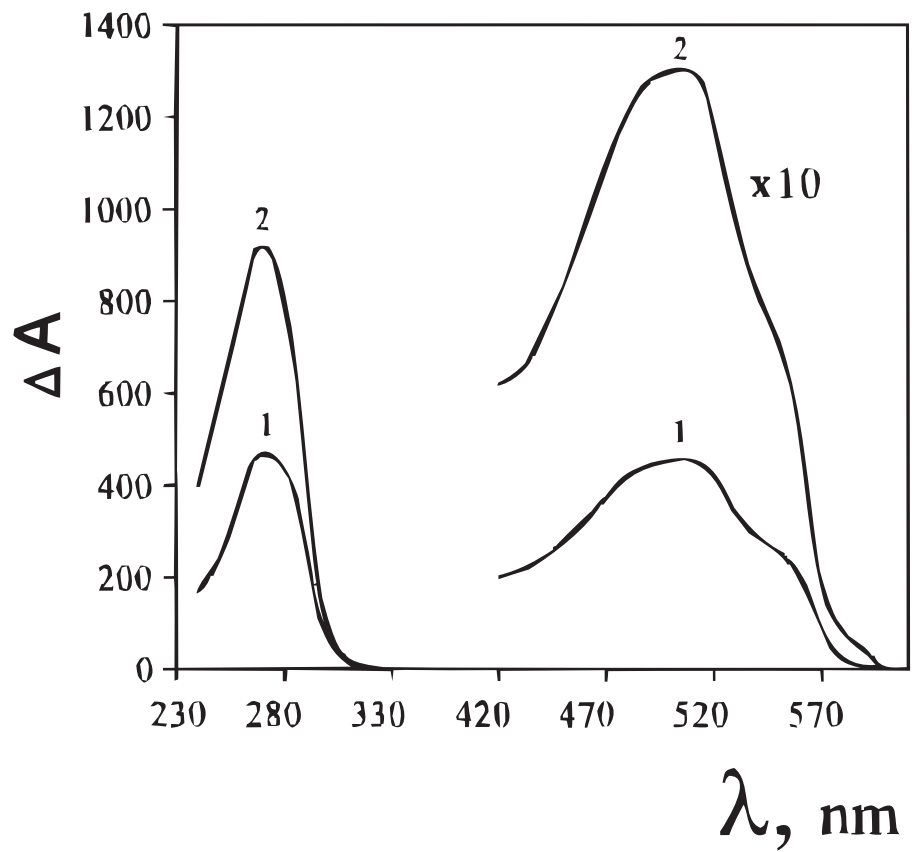
\end{document}